\documentclass[manuscript]{rmaa}

\usepackage{paralist}

\usepackage{psfrag,color}

\usepackage[latin1]{inputenc}
\usepackage{hyperref}
\title{Deep-TAO: The Deep Learning Transient Astronomical Object data set for Astronomical Transient Event Classification}

\author{
  John F. Su\'arez-P\'erez,\altaffilmark{1,2} 
  Catalina G\'omez,\altaffilmark{3}
  Mauricio Neira,\altaffilmark{4}
  Marcela Hern\'andez Hoyos,\altaffilmark{4}
  Pablo Arbel\'aez,\altaffilmark{5}, and
  Jaime E. Forero-Romero,\altaffilmark{2}
  }

\altaffiltext{1}{Tecnologico de Monterrey, Escuela de Ingenier\'ia y Ciencias, Zapopan, M\'exico (corresponding author: jf.suarez@tec.mx)}
\altaffiltext{2}{Departamento de F\'isica, Universidad de los Andes, Bogot\'a, Colombia}
\altaffiltext{3}{Department of Computer Science, Johns Hopkins University}
\altaffiltext{4}{Systems and Computing Engineering Department, Universidad de los Andes, Bogot\'a, Colombia}
\altaffiltext{5}{Center for Research and Formation in Artificial Intelligence, Universidad de los Andes, Bogot\'a, Colombia\\}

\shortauthor{Su\'arez-P\'erez, \& et al.}
\shorttitle{Deep-learning Transient Astronomical Object dataset}

\fulladdresses{
\item John F. Su\'arez P\'erez: Tecnologico de Monterrey, Escuela de Ingenier\'ia y Ciencias, Zapopan, M\'exico  (jf.suarez@tec.mx).

\item Catalina G\'omez: Department of Computer Science, Johns Hopkins University, Baltimore, MD, United States (cgomezc1@jhu.ed).
\item Mauricio Neira: Systems and Computing Engineering Department, Universidad de los Andes, Cra. 1 No. 18A-10, Bogot\'a, Colombia (m.neira@uniandes.edu.co)

\item Marcela Hern\'andez Hoyos: Systems and Computing Engineering Department, Universidad de los Andes, Cra. 1 No. 18A-10, Bogot\'a, Colombia  (marc-her@uniandes.edu.co)

\item Pablo Arbel\'aez: Center for Research and Formation in Artificial Intelligence, Universidad de los Andes, Cra. 1 No. 18A-10, Bogot\'a, Colombia (pa.arbelaez@uniandes.edu.co)

\item Jaime E. Forero-Romero: Departamento de F\'isica, Universidad de los Andes, Cra. 1 No. 18A-10, Bogot\'a, Colombia (je.forero@uniandes.edu.co)
}

\listofauthors{J. Su\'arez-P\'erez, C. G\'omez, M. Neira, M. Hern\'andez, P. Arbel\'aez, \& J. Forero-Romero}
\abstract{
We present the Deep-learning Transient Astronomical Object (Deep-TAO), a dataset of 1,249,079 annotated images from the Catalina Real-time Transient Survey, including 3,807 transient and 12,500 non-transient sequences. 
Deep-TAO has been curated to provide a clean, open-access, and user-friendly resource for benchmarking deep learning models. 
Deep-TAO covers transient classes such as blazars, active galactic nuclei, cataclysmic variables, supernovae, and events of indeterminate nature. 
The dataset is publicly available in FITS format, with Python routines and Jupyter notebooks for easy data manipulation. 
Using Deep-TAO, a baseline Convolutional Neural Network outperformed traditional random forest classifiers trained on light curves, demonstrating its potential for advancing transient classification.
}

\resumen{
Presentamos Deep-learning Transient Astronomical Object (Deep-TAO), un conjunto de 1,249,079 im\'agenes anotadas del Catalina Real-time Transient Survey, con 3,807 secuencias transientes y 12,500 no transientes.
Deep-TAO ha sido dise\~nado como un recurso limpio, de acceso abierto y f\'acil de usar, ideal para evaluar y comparar modelos de aprendizaje profundo.
Incluye eventos transientes como blazares, n\'ucleos gal\'acticos activos, variables catacl\'ismicas, supernovas y eventos de naturaleza indeterminada.
El conjunto de datos est\'a disponible p\'ublicamente en formato FITS, acompa\~nado de rutinas en Python y cuadernos de Jupyter que facilitan su uso.
Utilizando Deep-TAO, una red neuronal convolucional b\'asica super\'o el desempe\~no de clasificadores tradicionales basados en bosques aleatorios entrenados con curvas de luz, demostrando su potencial para mejorar la clasificaci\'on de eventos transientes.
}

\addkeyword{catalogues}
\addkeyword{transients: general}
\addkeyword{methods: data analysis}

\begin{document}
% Typeset article header
\maketitle

\section{Introduction} \label{sec:intro}
A primary challenge in time-domain astronomy lies in the detection and classification of transient astronomical events. 
Over recent years, methods to automate these processes have seen remarkable improvements in both complexity and computational efficiency, driven by the exponential growth of data sets requiring timely analysis \citep{Kaiser2004, Law2009, Smartt2015, Chambers2016, Martinez-Palomera2018, Bellm2019, Dyer2020, Nidever2021}.

Machine learning (ML) \citep{Wyrzykowski2014, Disanto2016, Gieseke2017, Neira2020, Sanchez-Saez2021, VanRoestel2021} and deep learning (DL) approaches \citep{Gieseke2017, Cabrera-Vives2017, Carrasco-Davis2019, Muthukrishna2019, Gomez2020, Sanchez-Saez2021, Allam2021, VanRoestel2021, Killestein2021} have demonstrated their capability to provide rapid and accurate solutions for transient classification tasks, offering significant advancements over traditional methods.

The further development and optimization of ML and DL algorithms critically depend on the availability of large-scale, high-quality, and representative data sets.
These data sets can be constructed from real observational data \citep{Neira2020}, synthesized light curves \citep{Carrasco-Davis2019}, or image-based data derived from either real \citep{Scalzo2017} or simulated observations \citep{Carrasco-Davis2019}.
The diversity and realism of these data sets are essential for improving the generalizability and robustness of classification models in the context of astronomical transient phenomena.

The image-based data sets that could be used to test and train new DL applications usually present some limitations:
\begin{itemize}
    \item[1)] Restricted access. Some data sets are private and only survey collaborators can access the data. 
    This limits the possibilities by a broader group of scientists to use the data set to improve DL techniques.
    \item[2)] Inconvenient access. Some surveys have setup public websites to access their data.
    However, sometimes the system has been designed to retrieve information about individual objects \citep{Drake2009, Scalzo2017, Nidever2021} and not large samples.
    This makes it inconvenient to compile the full data set required for DL training.
    \item[3)] Unrealistic images. 
    There are public and easy-to-gather data sets, however, they are based on simulated images \citep{Carrasco-Davis2019}. 
    This limits the realism desired to best train DL architectures.
    \item[4)] Incomplete labels. There are public, easy-to-gather, and realistic data sets that do not have labels on their data \citep{Smartt2015}. 
    These labels are required to train supervised DL architectures.
\end{itemize}

To date, no data set for DL transient classification has been made easily accessible to the public in the form of a fully labeled catalog based on observations. 
The purpose of this paper is to present a data set to fill that gap.

We denominate this data set Deep-TAO, for Deep-Learning Transient Astronomical Objects.
We build Deep-TAO using public data from the Catalina Real-Time Transient Survey (CRTS) \citep{Drake2009}, an astronomical survey searching for transient and highly variable objects.
We develop a procedure of extraction and transformation from CRTS into a homogeneous data set of thousands of objects that can be used to train DL algorithms and establish benchmarks.

\begin{figure*}
\begin{center}
   \includegraphics[width=1.0\linewidth]{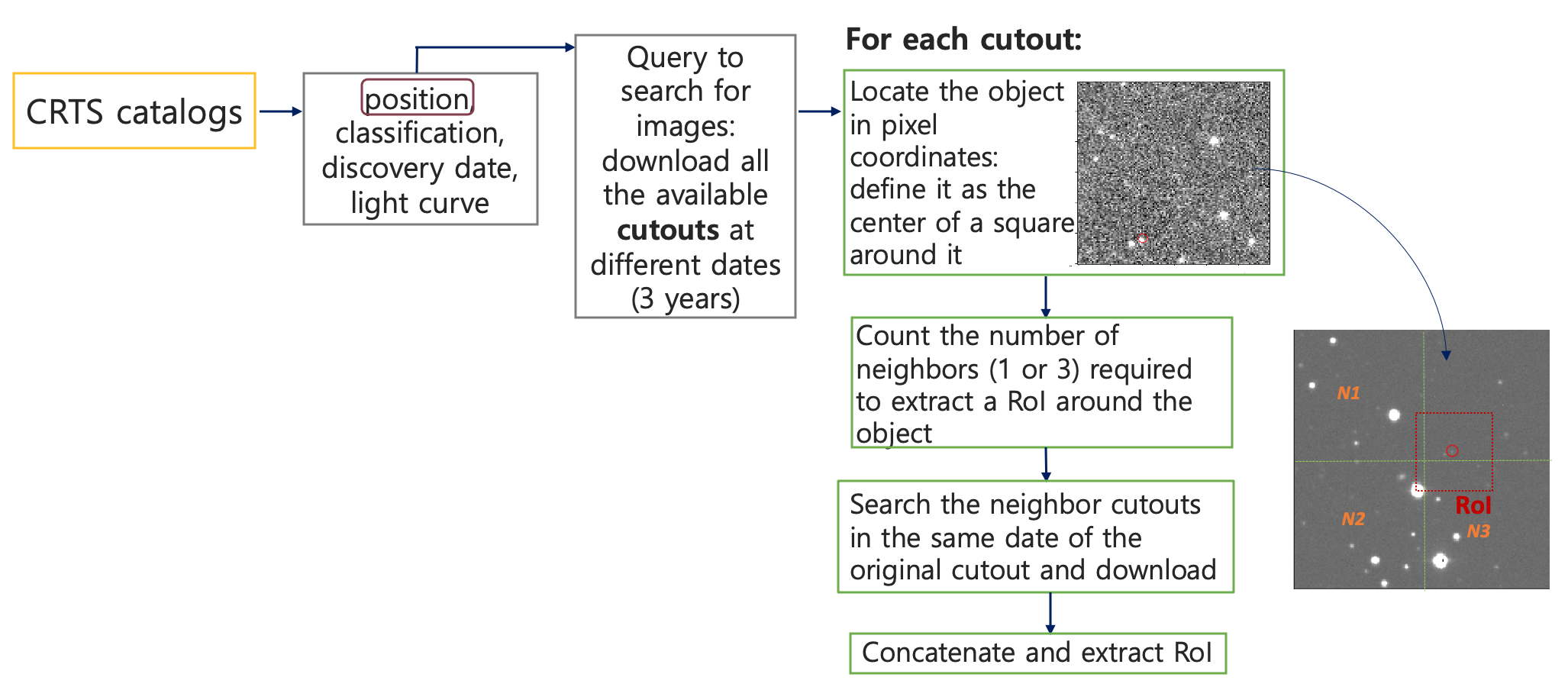}
\end{center}
   \caption{Overview of the search procedure to acquire the image sequences of transient objects.}
\label{fig:search_transients}
\end{figure*}

This paper is structured as follows. 
In Section \ref{sec:build} we describe the CRTS together with the selection and compilation procedures. 
In Section \ref{sec:features} we describe the main features of Deep-TAO including their structure.
Then, in section \ref{sec:mantra} we describe how to connect our data set with MANTRA \citep{Neira2020} a light curve-based dataset build also from the CRTS.
Finally, in section \ref{sec:methods} we demonstrate how Deep-TAO can be used in deep learning-based classification tasks, then we make a brief discussion and a summary.

\begin{figure*}
\begin{center}
   \includegraphics[width=\linewidth]{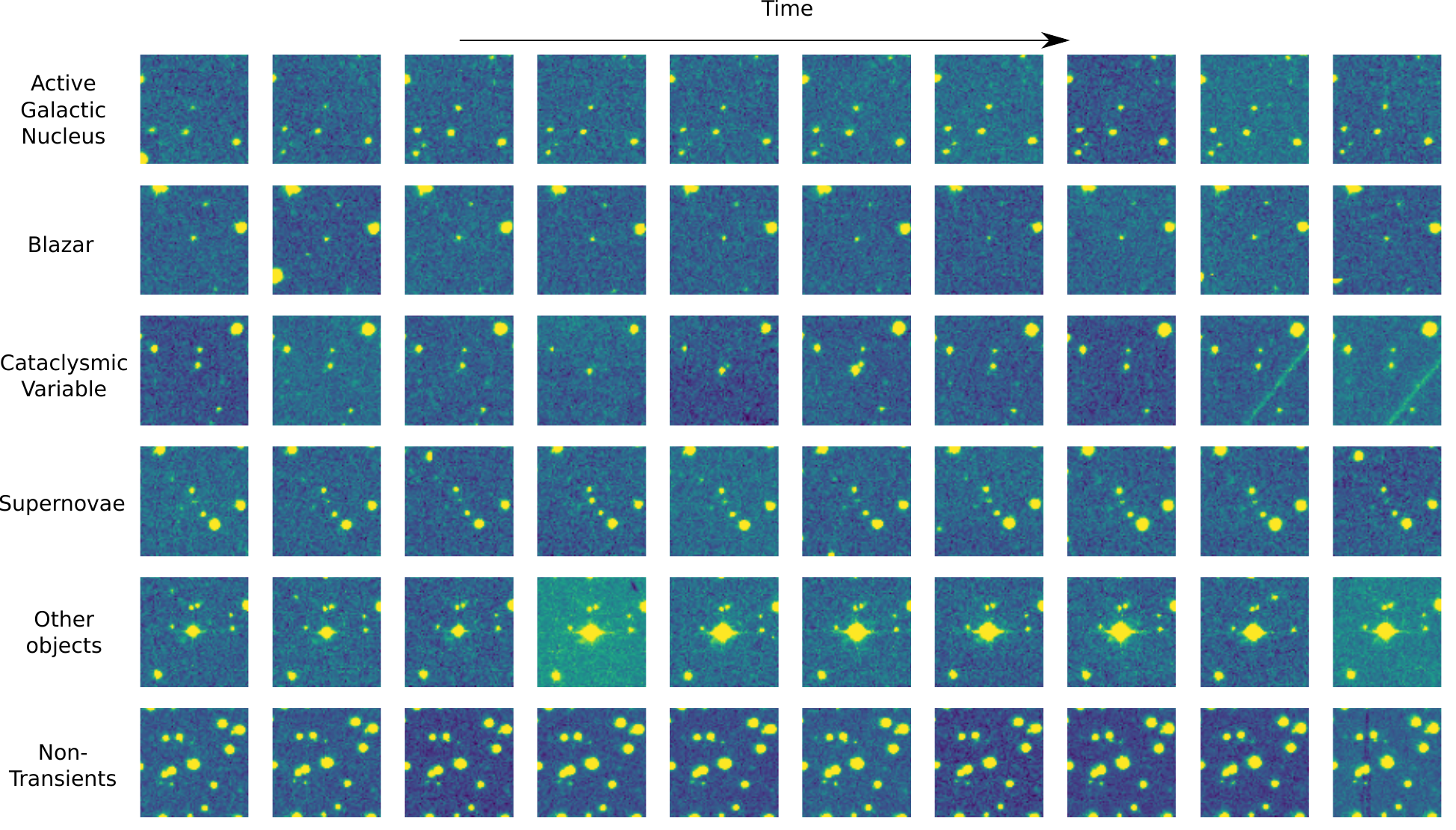}
\end{center}
\caption{Sample images in Deep-TAO. Each row corresponds to a sample of a different class. The temporal spacing between consecutive images varies for each example. Images were normalized for visualization.}
\label{fig:example}
\end{figure*}

\section{Observational Inputs to build Deep-TAO}\label{sec:build}

\subsection{The Catalina Real-Time Transient Survey and the Catalina Sky Survey }

We retrieve the images for Deep-TAO from the public catalogs of the Catalina Real-Time Transient Survey (CRTS) \citep{Drake2009, Mahabal2011}, an astronomical survey for transients and highly variable objects.
The area covered by the CRTS is 33,000 square degrees and has been observing the sky since 2007 with three telescopes: Mt. Lemmon Survey (MLS), Catalina Sky Survey (CSS), and Siding Spring Survey (SSS).
We use data from the CSS telescope, an $f/1.8$ Schmidt catadioptric equipped with a 111-megapixel CCD detector. 
The CSS telescope and detector have a scale of 2.5 arcseconds per pixel, giving an 8 square degrees field of view. 
Observations were made in a grid of adjacent fields. 
The survey covered 4,000 square degrees per night, with a limiting magnitude of 19.5 in the V-band.
Each observation is an image obtained using an exposure time of 30 seconds.

\subsection{Transient catalogs from the CRTS and the CSS}

We build Deep-TAO from the public transient catalog published by the CRTS.
The data reports five classes: blazars (BZ), active galactic nuclei (AGN), cataclysmic variables (CV), supernovae (SN), high proper motion stars (HPM), and other events of unknown nature \citep{Drake2009}.
The transient catalog lists the right ascension (RA), Declination (Dec), V-band magnitude, discovery date, classification class, and light curve points.

The CSS catalog contains observations from 2003 to 2012.
The selected fields were typically visited four times by night and the median total number of visits over 10 years is 20.
Each CSS image (of size $4,110 \times 4,096$ pixels covers an area of 29,500 squared arcminutes) is divided into 1,156 smaller images called cutouts stored in the Flexible Image Transport System (FITS) format. 
Each cutout is about $120 \times 120$ pixels and represents an area of $5 \times 5$ arcminutes.
Each cutout file stores the pixel intensities, the date on which the image was captured, a field identifier, and a number identifying the order of the image in the sequence of observations taken on a given night.

\subsection{Building Regions of Interest}
We use the cutouts to build a Region of Interest (RoI) centered on an object of interest.
We design RoIs to be squares of $64 \times 64$  pixels size centered on a RA/Dec coordinate of interest.
This requires downloading the cutouts, assembling them into a single image, and then finally cutting out the RoI around the RA/Dec of interest.

We refer to the time-ordered set of RoIs around the same coordinate as \emph{a RoI sequence}.
We build RoI sequences over a three-year interval, where the second year always includes the date of maximum brightness. 
Over that time the time-spacing between images is not uniform.
Intervals range from days to months.

We query the RoI sequences using web scraping techniques to automatically access and download the images using as an input a desired RA/Dec position.
This process comprises five different steps:

\begin{itemize}
    \item [1)] Download all the available cutouts that overlap with the input RA/Dec in a time span of three years for each object.
    \item [2)] For each cutout, locate the RA/Dec location to define a region of interest (RoI) around that coordinate.
    \item [3)] Count the number of neighboring cutouts (one or three) required to build the RoI. 
    \item[4)] Query for the neighboring cutouts. If any of those does not exist, the RoI is not built.
    \item[5)] Concatenate all cutouts to extract and store the RoI. 
\end{itemize}

Figure \ref{fig:search_transients} illustrate these steps.
It took $11,000$ CPU hours to query the CRTS/CSS database to build the full Deep-TAO data set.

The transient objects are available from CRTS catalogs.
However, a catalog of Non-Transient objects is not available.
To define Non-Transient RA/Dec locations, we use the transient sources cutouts. 
All sources in the cutout of a transient at any date are detected. 
Then, the sources at a distance greater than a threshold of 33 pixels from the transient are considered as a possible non-transient candidate. 
This threshold ensures that the transient object does not appear in the RoI of the non-transient candidate.
For each one of the non-transient candidates we compute its RA/Dec coordinates to build all the RoIs on the same dates as the parent transient sequence.
Using this procedure, we compile a total of 12,500 Non-Transient locations.

\section{Deep-TAO Description}\label{sec:features}
Figure \ref{fig:example} shows a grid of illustrative examples for different transients and Non-Transients in Deep-TAO.
The images in that figure are a subset from the full RoI sequence for each object,
the temporal spacing between images is uneven and the time-stamps are not uniform across different objects.
To ease visualization the pixel values are re-normalized to have the same range across all the images.

In all the cases shown in Figure \ref{fig:example}, the variability of the central source is easy to spot by eye.
This illustrative example also shows features (i.e. trails at the end of the Cataclysmic Variable sequence, overall brightness change in the first half of the Other Objects class) that might come from fluctuating observational and instrumental conditions, representing the realism of Deep-TAO.

Out what follows we describe overall Deep-TAO statistics, the data model used to store the information in the public repositories, and the python-based tools to interact with Deep-TAO files.

\begin{figure*}
\begin{center}
   \includegraphics[width=0.46\linewidth]{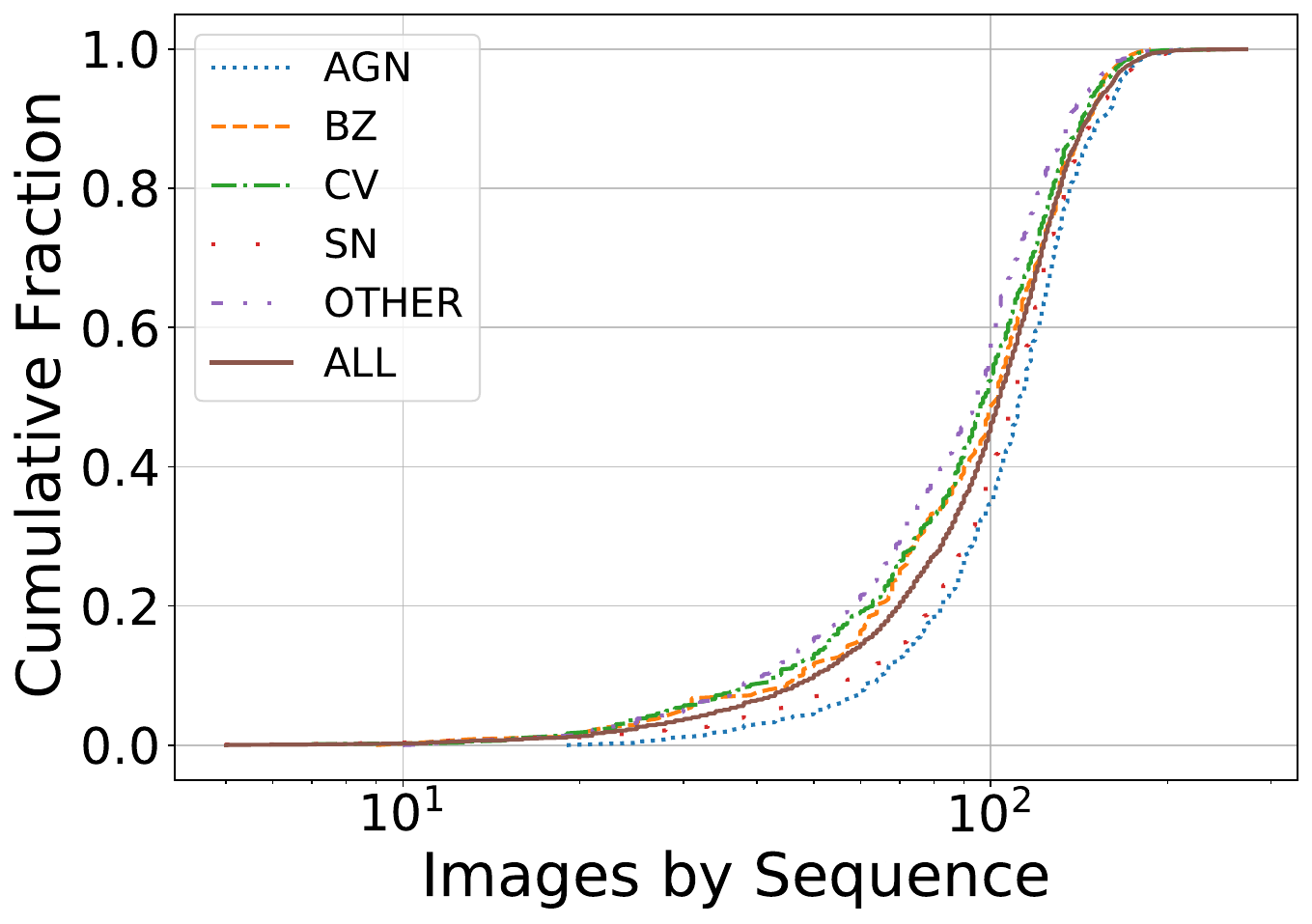}
   \includegraphics[width=0.46\linewidth]{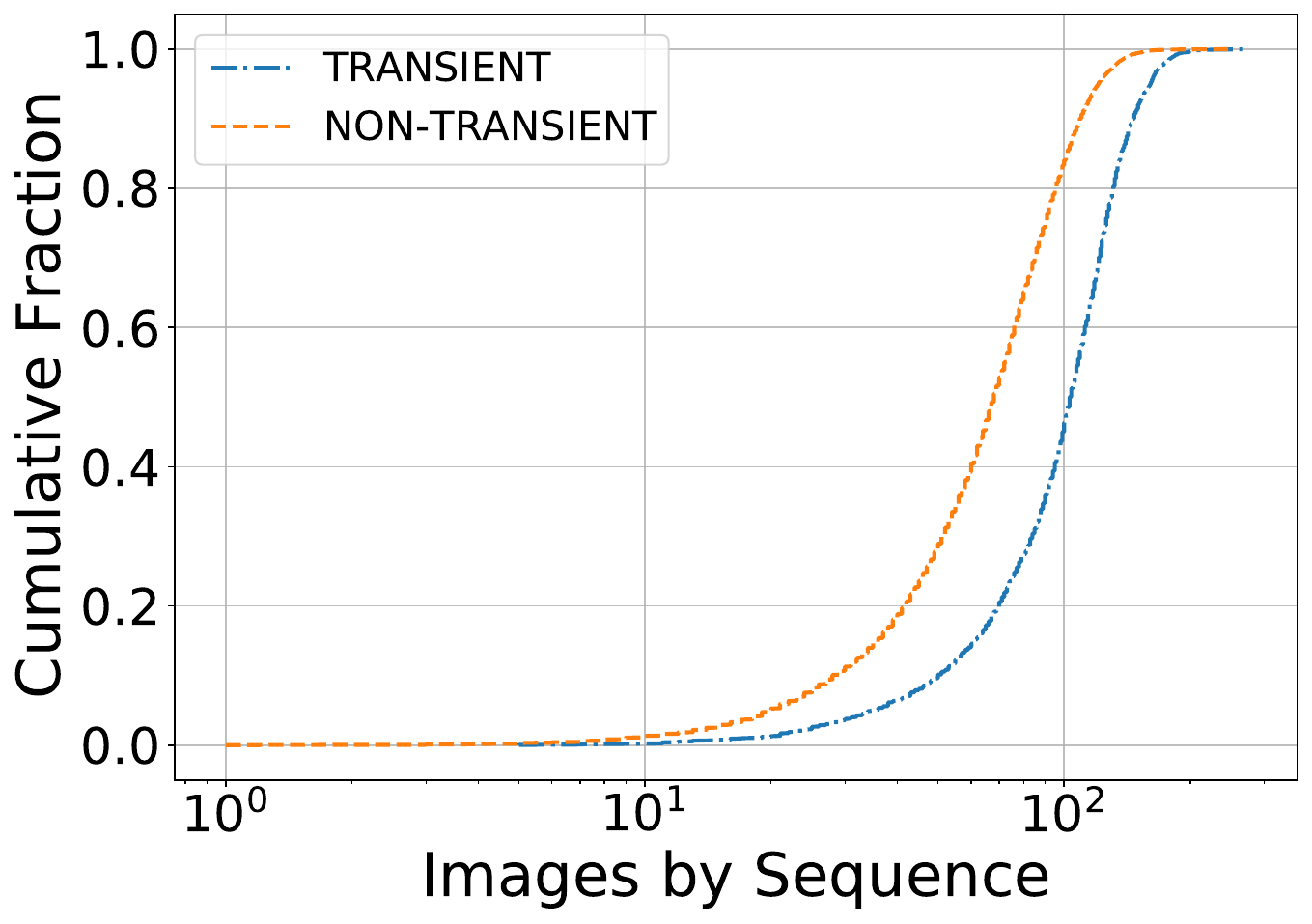}
\end{center}
   \caption{Cumulative distribution of RoIs per sequence. (Left) the distributions are split across transient classes. The median is around 100 images by sequence. (Right) Distributions split between transients and non-transients objects. The median for non-transients is around 70 images per sequence.}
\label{fig:cumulative_seqs}
\end{figure*}

\begin{figure*}
\begin{center}
   \includegraphics[width=0.46\linewidth]{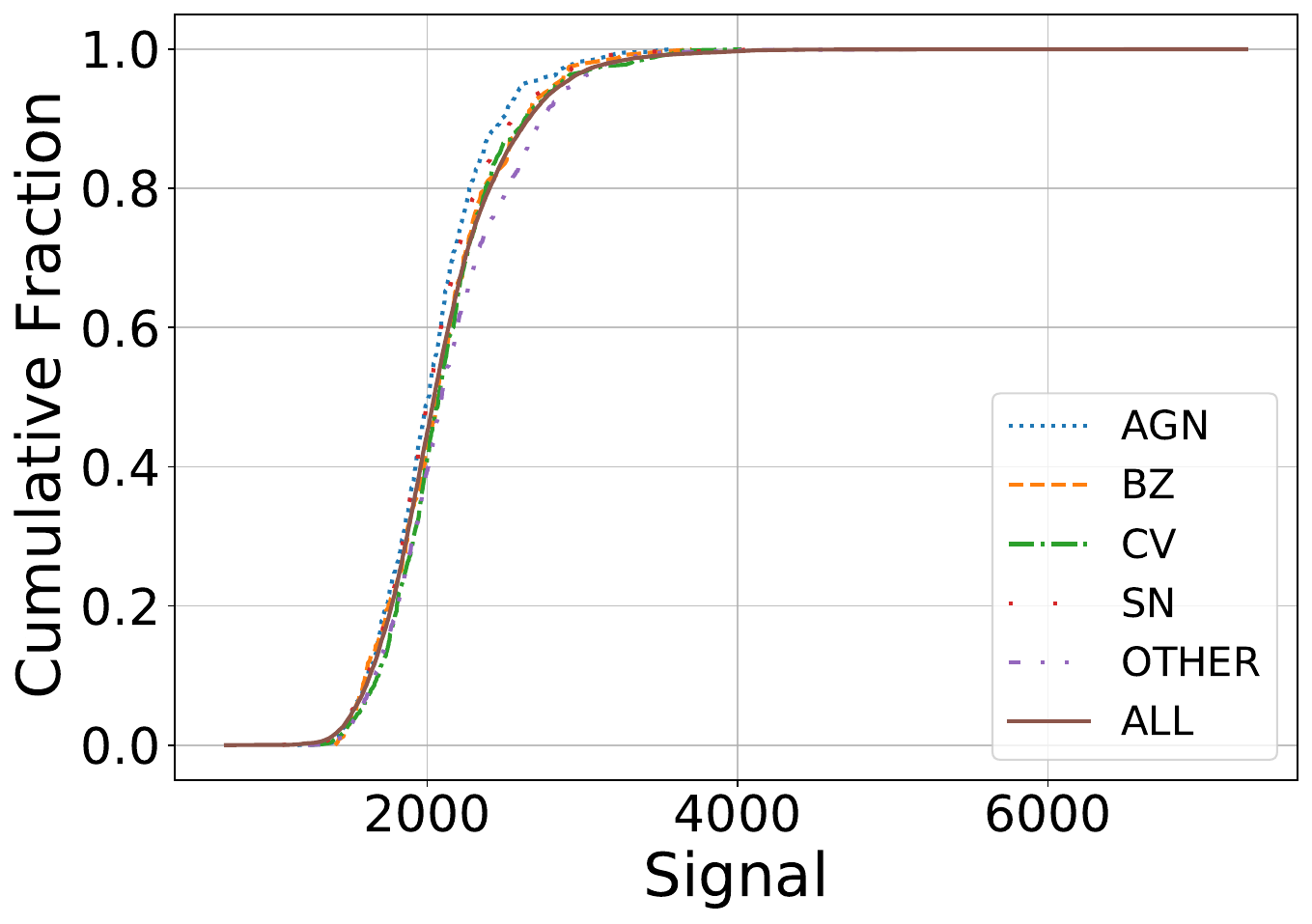}
   \includegraphics[width=0.46\linewidth]{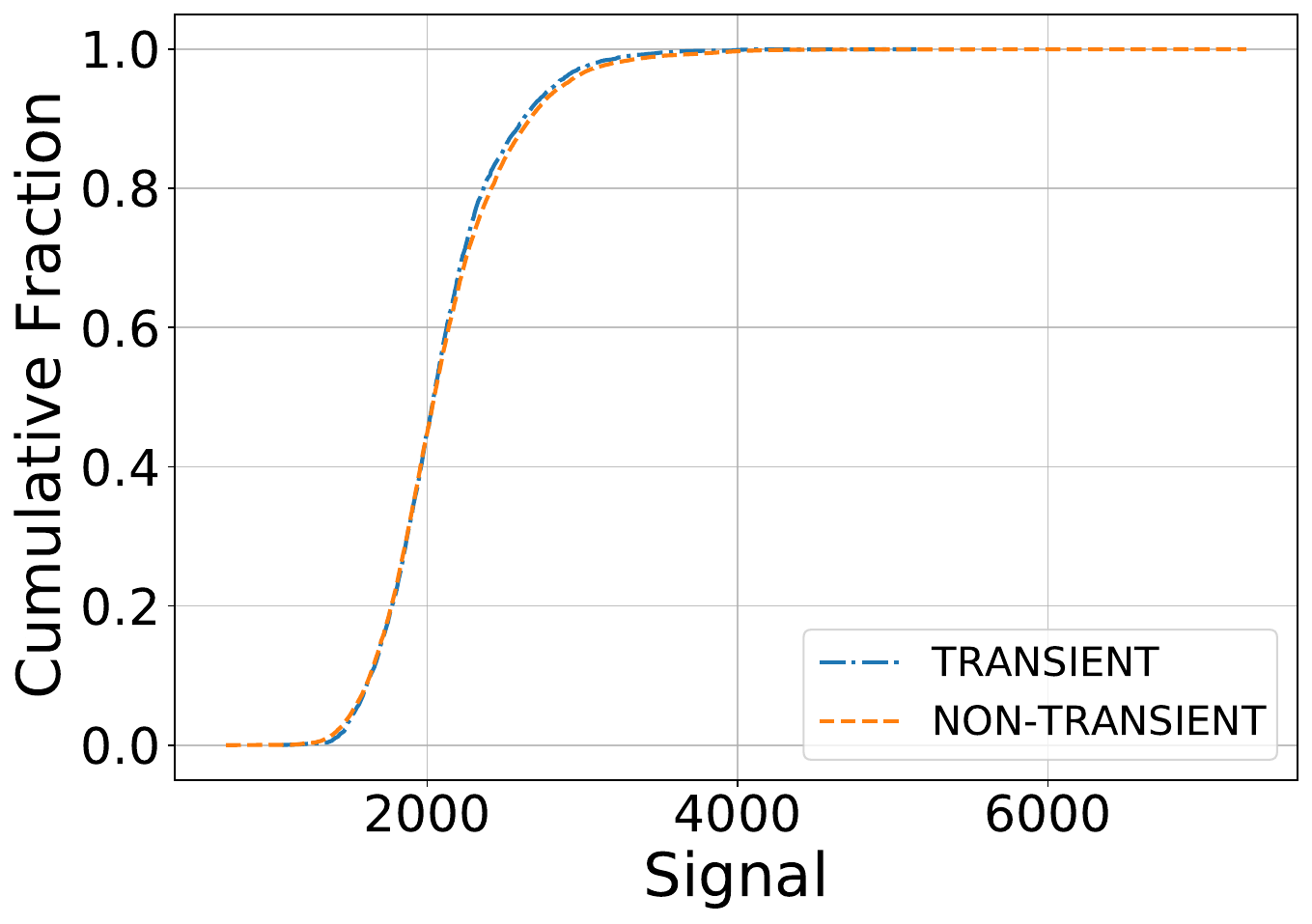}
\end{center}
   \caption{(Left) Cumulative fraction as a function of the median signal for the objects in each transient class and all transients objects (continuous line). (Right) Cumulative fraction between transients and non-transients objects. The shape between these classes is similar. In both Figures the media of the signal is around 2000.}
\label{fig:cumulative_signal}
\end{figure*}

\begin{figure*}
\begin{center}
   \includegraphics[width=0.45\linewidth]{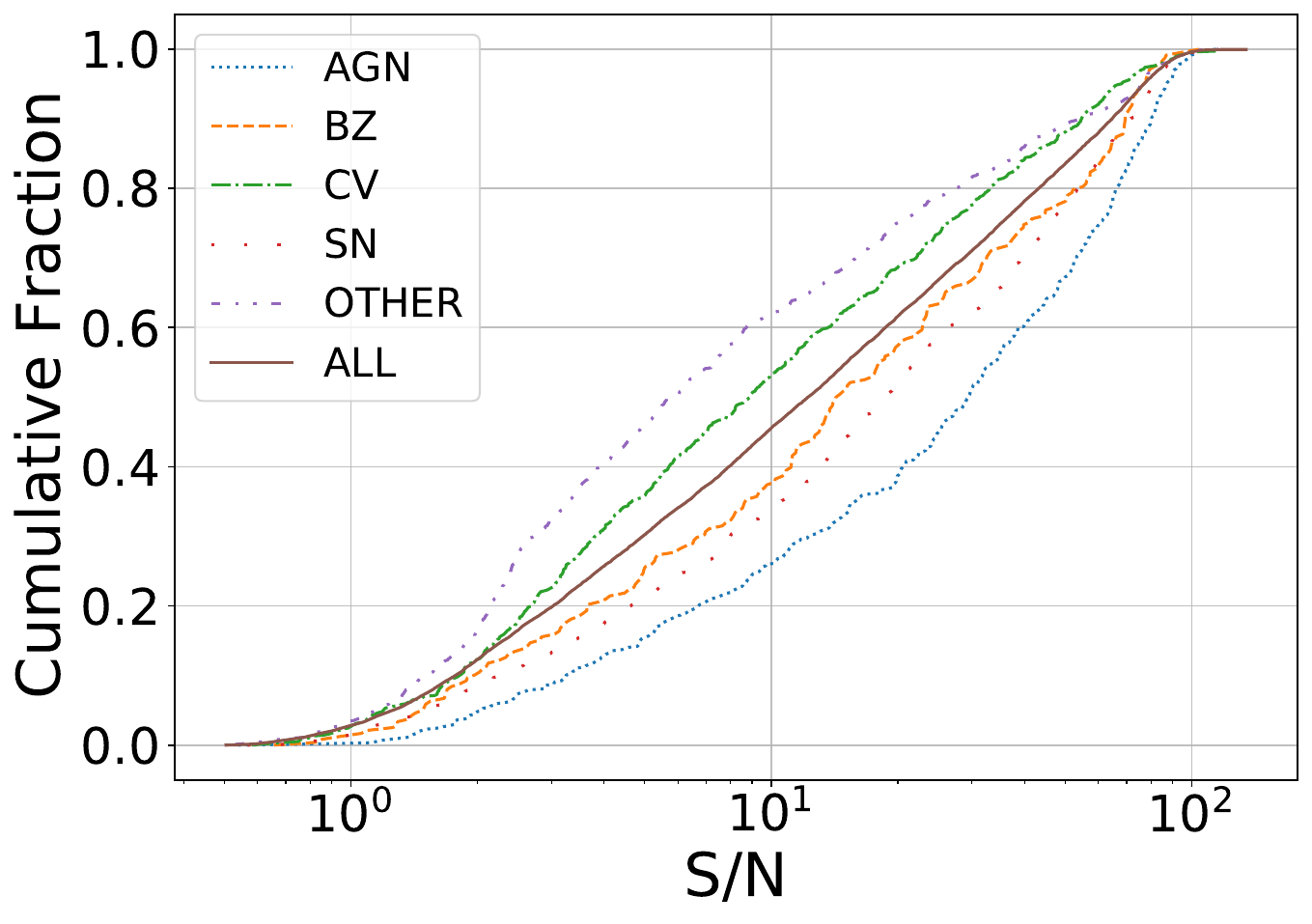}
   \includegraphics[width=0.45\linewidth]{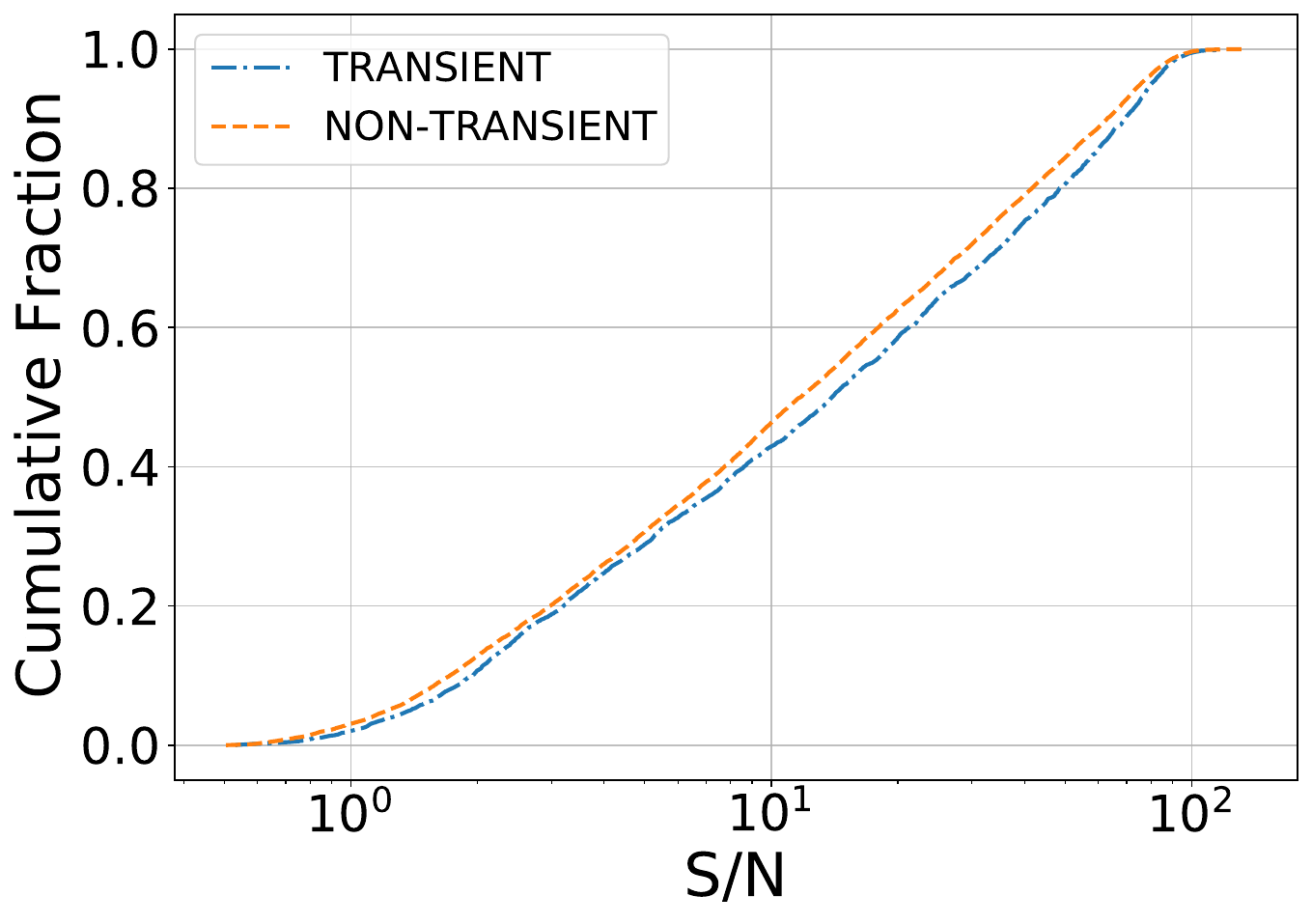}
\end{center}
   \caption{(Left) Cumulative distribution of the average signal-to-noise. for the object in each transient class and all transients objects (continuous line). The media of signal/noise for all transients objects is around 10. (Right) Cumulative fraction between transients and non-transients objects. In both cases the media of signal/noise is around 20.}
\label{fig:cumulative_signalnoise}
\end{figure*}

\subsection{General Statistics}

Table \ref{tab:stats} summarizes global statistics for the Deep-TAO data set.
The first row shows the total number of targets in the original CRTS catalog.
The second row indicates the number of targets for which we manage to recover a RoI sequence.
Some transients in the original catalog are not included in Deep-TAO out due to the impossibility of having the transient centered in the cutout.
The third row indicates the total number of RoI extracted for each class.

\begin{table}[!t]\centering
  \setlength{\tabnotewidth}{1.5\columnwidth}
  \tablecols{9}
  \begin{changemargin}{-3.5cm}{-3.5cm}
  % Stretch the space between table columns
  %\setlength{\tabcolsep}{2.8\tabcolsep}
  \caption{General statistics of Deep-TAO data set\tabnotemark{*}} \label{tab:stats}
  \begin{tabular}{ccccccccc}
    \toprule
    & \multicolumn{1}{c}{BZ} & \multicolumn{1}{c}{AGN} & \multicolumn{1}{c}{CV} & \multicolumn{1}{c}{OTHER} & \multicolumn{1}{c}{SN} & \multicolumn{1}{c}{Total Transients} & \multicolumn{1}{c}{Non-Transients} & \multicolumn{1}{c}{Total} \\
    \midrule
    Targets in CRTS    & 270 & 651 & 987 & 1,054 & 1,723 & 4,712 & -      & 4,712 \\
    Targets in Deep-TAO & 239 & 606 & 772 & 818   & 1,372 & 3,807 & 12,500 & 16,307 \\
    Total RoIs         & 23,429 & 66,998 & 73,739 & 74,536 & 146,847 & 385,549 & 863,530 & 1,249,079 \\
    \bottomrule
    \tabnotetext{a}{The first row corresponds to the transients included in the public CRTS transient catalog. The second row represents the number of objects for which we can retrieve a sequence of RoIs over a three-year observation period. The last row is the total number of RoIs included for each class.}
  \end{tabular}
  \end{changemargin}
\end{table}

\begin{table}[!t]\centering
  \setlength{\tabnotewidth}{1\columnwidth}
  \begin{changemargin}{-1.5cm}{-1.5cm}
  \tablecols{3}
  % Stretch the space between table columns
  \setlength{\tabcolsep}{1\tabcolsep}
  \caption{FITS header of the transient files} \label{tab:header_trans}
  \begin{tabular}{ccc}
    \toprule
    Header Dict & Description & Type \\
    \midrule
    CRTS\_ID             & Catalina Real-time Transient Survey ID. & str \\
    RA\_(J2000)          & Right Ascension (degrees). & float \\
    Dec\_(J2000)         & Declination (degrees). & float \\
    N\_Images            & Total number of images for CRTS ID. & int \\
    UT\_Date             & UT Discovery Date (YYYYMMDD). & float \\
    Mag                  & Unfiltered CSS magnitude. & float \\
    CSS\_Images          & Pre and post-discovery images ID. & int \\
    SDSS                 & Covered by SDSS DR-12 (yes/no). & str \\
    Others               & ID to other image data at the location (PQ, DSS, 2MASS, SDSS). & int \\
    Followed             & P60 follow up (yes/no). & str \\
    Last                 & Last Observation date. & str \\
    LC                   & Current CSS lightcurve. & int \\
    FC                   & Finding chart (yes/no). & str \\
    Class                & Transient classification. & str \\
    \bottomrule
  \end{tabular}
  \end{changemargin}
\end{table}

\begin{table}[!t]\centering
  \begin{changemargin}{-3.5cm}{-3.5cm}
  \setlength{\tabnotewidth}{1.5\columnwidth}
  \tablecols{3}
  % Stretch the space between table columns
  \setlength{\tabcolsep}{1\tabcolsep}
  \caption{Identifiers stored first HDU: Transient files\tabnotemark{*}} \label{tab:hdu_trans}
  \begin{tabular}{ccc}
    \toprule
    \textbf{Key} & \textbf{Description} & \textbf{Type} \\
    \midrule
    HDU\_Ext     & HDU extension of the RoI (From 2 to \texttt{N\_Images+1}). & int \\
    Set\_Number  & Stands for the sequence (or set number). & str \\
    Date         & Date of observation (YYMMMDD). & str \\
    MJD          & Modified Julian Date. & float \\
    Field\_ID    & Field identifier. & str \\
    Obs\_In\_Seq & Refers to the observation's number in the sequence. & str \\
    Cutout       & The cutout matrix location. Each cutout covers an area of ABOUT 5 x 5 arcminutes. & str \\
    \bottomrule
    \tabnotetext{a}{Basic information in the first HDU about the image sequence in each transient FITS file.}
  \end{tabular}
  \end{changemargin}
\end{table}

\begin{table}[!t]\centering
  \begin{changemargin}{-1.5cm}{-1.5cm}
  \setlength{\tabnotewidth}{0.5\columnwidth}
  \tablecols{3}
  % Stretch the space between table columns
  \setlength{\tabcolsep}{1\tabcolsep}
  \caption{FITS header of the non-transient objects.} \label{tab:header_non}
  \begin{tabular}{ccc}
    \toprule
    \textbf{Header Dict} & \textbf{Description} & \textbf{Type} \\
    \midrule
    CRTS\_ID   & Catalina Real-time Transient Survey ID. & str \\
    RA\_(J2000) & Right Ascension (degrees). & float \\
    Dec\_(J2000) & Declination (degrees). & float \\
    N\_Images & Total number of images for CRTS ID. & int \\
    Img\_Ref & Image of reference where the non-transient object was identified. & str \\
    \bottomrule
  \end{tabular}
  \end{changemargin}
\end{table}

\begin{table}[!t]\centering
  \begin{changemargin}{-3cm}{-3cm}
  \setlength{\tabnotewidth}{1.5\columnwidth}
  \tablecols{3}
  % Stretch the space between table columns
  \setlength{\tabcolsep}{1\tabcolsep}
  \caption{Identifiers stored first HDU: Transient files\tabnotemark{*}} \label{tab:hdu_non}
  \begin{tabular}{ccc}
    \toprule
    \textbf{Key} & \textbf{Description} & \textbf{Type} \\
    \midrule
    HDU\_Ext     & HDU extension of the RoI (From 2 to \texttt{N\_Images+1}). & int \\
    Date         & Date of observation (YYMMMDD). & str \\
    MJD          & Modified Julian Date. & float \\
    Field\_ID    & Field identifier. & str \\
    Cutout       & The cutout matrix location. Each cutout covers an area of ABOUT 5 x 5 arcminutes. & str \\
    \bottomrule
    \tabnotetext{a}{Basic information in the first HDU of the Non-Transient objects about the image sequence in each FITS file.}
  \end{tabular}
  \end{changemargin}
\end{table}

Figures \ref{fig:cumulative_seqs}, \ref{fig:cumulative_signal}, \ref{fig:cumulative_signalnoise} present some cumulative statistics computed over the RoI sequences for each class.
Figure \ref{fig:cumulative_seqs} shows the cumulative distribution for the number of images by sequence. 
The left panel shows all the transient classes, the right panel compares transients and non-transients.
This Figure shows that the median value is close to 100 RoIs per sequence. 
The shortest sequence has 5 RoIs and the longest close to 300 RoIs.
For non-transient sequences, there is a median of 70 RoIs, while for transients the median is 100 RoIs per sequence.

Figure \ref{fig:cumulative_signal} shows results for the average RoI signal.
Here we define the signal as the sum of all CCD counts across the RoI.
The left panel corresponds to all transient classes, while the right panel compares transients and non-transients.
This Figure shows that all transient classes and the non-transients have similar intensity distributions.

Figure \ref{fig:cumulative_signalnoise} compares the average signal-to-noise (S/N) distribution for all transient classes (left) and transients versus non-transients (right).
We estimate the signal-to-noise for a RoI as the ratio between the sum of all CCD counts and the standard deviation of the CCD counts.
We find that the average S/N spans almost two orders of magnitude ranging from 1 up to 100. 
For transients, the median of the average S/N ranges between 6 and 20 across all classes, with some differences between classes.
On the contrary, the distribution for Transients and Non-Transients is virtually the same.

\subsection{Data Model}
The Deep-TAO data set is allocated on GitHub into two different repositories, one for transients objects\footnote{ \url{https://github.com/MachineLearningUniandes/TAO_transients}} and other for non-transients\footnote{\url{https://github.com/MachineLearningUniandes/TAO_non-transients}}. The transient's repository contains three main folders \texttt{data}, \texttt{paper} and \texttt{mantra}.

The \texttt{data} folder has all the transients sequences separated in subfolders by class (AGN, BZ, CV, OTHERS, and SN), each subfolder contains the sequences stored in FITS files.
A single FITS file stores all the RoIs associated with a transient event, the file name is the CRTS identifier. 
Each file contains a header, the FITS header in each file has minimal identifying information such as the CRTS\_ID unique identifier, the J2000 RA/Dec coordinates, the number of RoIs (\texttt{N\_Images}) in the sequence, and the Universal Time UT\_Date associated with the discovery date.
The full list of fields included in the header is listed in Table \ref{tab:header_trans}.

The first HDU (extension 1) in the FITS files is a 2D array with the columns listed on Table \ref{tab:hdu_trans}.
This array contains information for each RoI in the sequence, such as the HDU extension for each RoI and its observation date. 
Starting from the HDU 2 on-wards up to the HDU \texttt{N\_Images+1}, each HDU contains a RoI as an integer array of size 64$\times$64.

The second main folder is \texttt{paper}, it contains the Figures \ref{fig:cumulative_seqs}, \ref{fig:cumulative_signal},  \ref{fig:cumulative_signalnoise} that describe general statistics of Deep-TAO. 
This folder also contain a python-based tool to reproduce these ones. 
This tool will be explained in the next subsection.

The \texttt{mantra} folder contains the Figure \ref{fig:TAO_MANTRA}, which shows an example of how to connect Deep-TAO with MANTRA (Many ANnotated TRAnsients), an annotated Machine-learning Reference lightcurve data set in V-band also built from the CRTS \citep{Neira2020}. 
More details in the section \ref{sec:mantra}.

Finally, the non-transient's repository only contains the \texttt{data} folder with the FITS files of the non-transients objects. 
At difference to the header of a Transient FITS file, the non-transient FITS header allocate the information of the Table \ref{tab:header_non}, these are the CRTS\_ID, the RA/Dec coordinates, the number of images in the sequence, and the image source where was extracted from.

The first HDU in each FITS non-transient file allocates the information of the Table \ref{tab:hdu_non}: the HDU\_Extension for each RoI, the date of observation, the MJD, the Field\_ID, and the cutout. 

\subsection{Python-based tools}

In the folder \texttt{data} of the transients repository, there is a jupyter notebook to manipulate the data. The \texttt{Read\_dataset} jupyter notebook shows the mechanism to read the FITS files for the transients and non-transients objects. 
In the folder \texttt{paper} in the same repository, we provide the \texttt{Explore\_data set} Jupyter notebook, this shows how to compute some statistics from Deep-TAO to obtain the Figures \ref{fig:example}, \ref{fig:cumulative_seqs}, \ref{fig:cumulative_signal} and \ref{fig:cumulative_signalnoise}, assuming that the \texttt{data/NON} folder from the non-transient's repository is located in the \texttt{data} folder of the transient's repository.
This notebook also create a plain text file in the \texttt{paper} folder called \texttt{statistic.csv}. 
This file has 16,307 rows, one by object in Deep-TAO, and four columns with the class name \texttt{class} (BZ,AGN,CV,OTHER,SN or NON), the number of images by sequence \texttt{nimages\_seq}, the median of the signal/noise measure \texttt{signal\_noise\_median} and the median of the signal \texttt{signal\_median}.

\section{Linking Deep-TAO images to MANTRA lightcurves}\label{sec:mantra}

\begin{figure*}
\centering
    \includegraphics[width=1\linewidth]{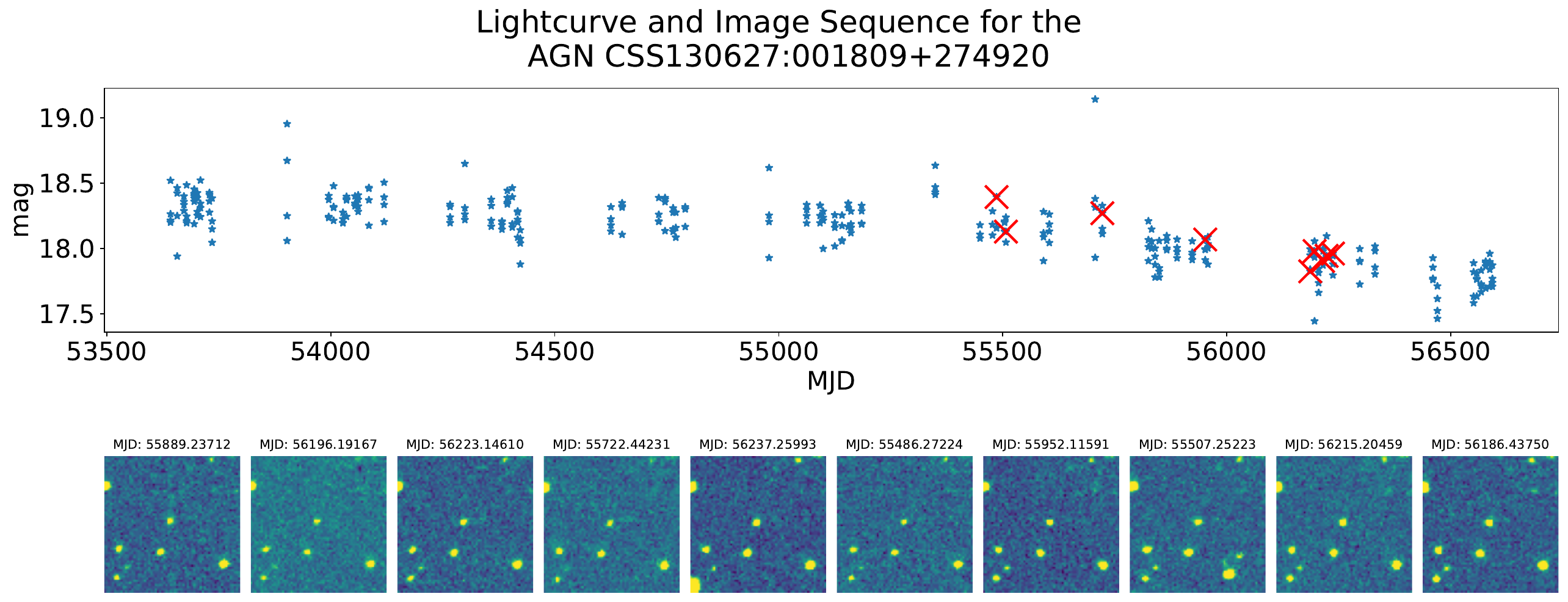}
\caption{Lightcurve and examples of the image sequence for the AGN CSS130627:001809+274920 
from MANTRA and Deep-TAO obtained using the \texttt{Connection\_MANTRA} jupyter notebook. The red cross correspond to the images plotted bottom in the Figure.}
\label{fig:TAO_MANTRA}
\end{figure*}

In \citep{Neira2020} was presented MANTRA an annotated Machine-learning reference lightcurve data set also built from the CRTS. 
MANTRA contains 4,869 transients and 71,207 Non-Transients as a plain text file to facilitate standardized quantitative comparison of astronomical transient event recognition algorithms. 
The classes included in MANTRA are Supernovae, Cataclysm Variable, Active Galactic Nuclei, High Proper Motion stars, Blazars, and Flare. 
The data set is publicly available and easy to access \footnote{\url{https://github.com/MachineLearningUniandes/MANTRA}}. 

In the \texttt{mantra} folder of the Deep-TAO transients repository\footnote{ \url{https://github.com/MachineLearningUniandes/TAO_transients}}, we provide the \texttt{Connection\_MANTRA} Jupyter notebook to link the image sequence from Deep-TAO to the lightcurve from MANTRA. 
This connection is done through the unique CRTS ID.
For non-transients, this connection between images and light curves cannot be established  between Deep-TAO and MANTRA due both have different non-transient objects.

Figure \ref{fig:TAO_MANTRA} shows an example for an AGN.
Using the MJD information it is possible to connect points in the light curve to images in the sequence. 
In the light curve of Figure \ref{fig:TAO_MANTRA}, the red crosses correspond to the images plotted below in the Figure. 
Due to the constraints in the RoI construction (Section \ref{sec:build}) not all points in the MANTRA lightcurve have a corresponding image in Deep-TAO. 
Also, because Deep-TAO includes only three-year intervals of observations.

\section{Example of a Deep-TAO application}\label{sec:methods}

Here we show some examples of Deep-TAO applications using a Convolutional Neural Network (CNN) to gauge its performance on three basic classification tasks:
\begin{enumerate}
\item binary classification between Transients and Non-Transients.
\item fine-grained classification into five transient classes (Blazar, AGN, Cataclysmic Variables, Supernovae, and Other)
\item fine-grained classification into five transient classes and Non-Transients as a sixth class.
\end{enumerate}

We evaluate all tasks with metrics robust to class imbalance. 
For each class, we report the maximum F-measure (F1) from the Precision-Recall (PR) curve that we construct by setting different thresholds on the output probabilities of each class. 
The global performance is the F1 average across individual classes with an uncertainty computed as the standard deviation.
In all the experiments we use 70\% of Deep-TAO for training, 25\% for validation, and 5\% for testing.

The CNN we use here is based on previous work by \citet{Gomez2020}.
They used TAO-Net a neural network composed of two modules.
First, a CNN based on the architecture DenseNet to extract a feature representation and then a Recurrent Neural Network (RNN) that uses these representations to solve the classification task.
Here we only use the first part, a CNN based on a Densely Connected Convolutional Network (DenseNet) \citep{Huang2017} with $L=70$ layers and a growth rate $k=32$.

We model the temporal information by selecting images from the complete sequences. 
We consider images at three different dates in sequential order, such that they reflect differences in brightness for transient classes. 
We include the observation date in the three-year period when the transient object had the maximum brightness and one observation before and after that date. 
For the Non-Transient class, we take the first, middle, and last dates of the sequence of ordered images. 
At each date, we take the first available observation, and then merge the temporal information by sampling images from the complete sequences at three different dates in sequential order.
That selection reflects the evolution of the temporal information evidencing the differences in the brightness for transient classes.

Table \ref{tab:binary} summarizes the results of the binary classification task.
As expected, it is considerably easier to classify a sequence as a Non-Transient (F1 of 95.06) than as Transient (F1 of 74.46).

For the five-class transient classification task, we made a experiment that consist in the traditional approach for transient classification using the light curves from the CRTS. We compute the discriminatory features from the light curves to train a Random Forest (RF) classifier. 
All the details on the feature extraction and the RF classifier can be found in \cite{Neira2020}. 
These results are equal to \citet{Gomez2020} because we share the same dataset and the algorithm parameters.

\begin{table}[!t]\centering
  \setlength{\tabnotewidth}{1\columnwidth}
  \tablecols{9}
  \tablecols{6}
  % Stretch the space between table columns
  \setlength{\tabcolsep}{0.5\tabcolsep}
  \caption{F-measure for the binary task\tabnotemark{*}} \label{tab:binary}
  \begin{tabular}{cccccc}
    \toprule
    \textbf{Set} & \textbf{Data} & \textbf{Model} & \textbf{Transient} & \textbf{Non-Transient} & \textbf{F1} ($\mu \pm \sigma$) \\
    \midrule
    Validation & Images & TAO-Net & 74.46 & 95.06 & 84.76 $\pm$ 10.30 \\
    % Test & Images  &  TAO-Net & 88.38 & 96.62 & $92.50 \pm 4.12$ \\ 
    \bottomrule
    \tabnotetext{a}{F-measure for each class in the validation set for the binary task. The last column reports the average F-measure.}
  \end{tabular}
\end{table}

\begin{table}[!t]\centering
  \setlength{\tabnotewidth}{1\columnwidth}
  \tablecols{9}
  % Stretch the space between table columns
  \setlength{\tabcolsep}{0.5\tabcolsep}
  \caption{F-measure for the transient classification\tabnotemark{*}} \label{tab:transients}
  \begin{tabular}{ccccccccc}
    \toprule
    \textbf{Set} & \textbf{Data} & \textbf{Model} & \textbf{BZ} & \textbf{AGN} & \textbf{CV} & \textbf{OTHER} & \textbf{SN} & \textbf{F1} ($\mu \pm \sigma$)\\
    \midrule
    Validation & Light curves & RF & 19.74 & 42.67 & 53.60 & 56.06 & 55.36 & 45.49 $\pm$ 13.75 \\
    Validation & Images & CNN & 25.17 & 49.77 & 59.48 & 64.04 & 63.39 & 52.37 $\pm$ 14.53 \\
    \bottomrule
    \tabnotetext{a}{The last column reports the average F-measure of the 5 transient categories.}
  \end{tabular}
\end{table}

\begin{table}[!t]\centering
  \setlength{\tabnotewidth}{1\columnwidth}
  \tablecols{10}
  % Stretch the space between table columns
  \setlength{\tabcolsep}{0.5\tabcolsep}
  \caption{F-measure for the multi-class detection\tabnotemark{*}} \label{tab:results_multi}
  \begin{tabular}{cccccccccc}
    \toprule
    \textbf{Set} & \textbf{Data} & \textbf{Model} & \textbf{BZ} & \textbf{AGN} & \textbf{CV} & \textbf{OTHER} & \textbf{SN} & \textbf{Non-T} & \textbf{F1} ($\mu \pm \sigma$) \\
    \midrule
    Validation & Images & CNN & 21.82 & 37.45 & 54.76 & 40.22 & 46.59 & 95.29 & 49.36 $\pm$ 22.84 \\
    \bottomrule
    \tabnotetext{a}{The last column reports the average F-measure of the 6 classes.}
  \end{tabular}
\end{table}

Table \ref{tab:transients} shows the F-scores of the transient classification tasks.
The results show that classification with images using a CNN is a better option that makes a classification with light curves using a RF algorithm. 
With RF on light curves, the best classification is for the OTHER class with 56.06, followed by the SN class with 55.36. The worst is the BZ class with 19.74, the average F1-score is 45.49. The CNN on images is better with an average F1-score of 52.37, where the best classification is for OTHER with 64.04, followed by SN with 63.39 and the worst classification is for BZ with 25.17.

Finally, in Table \ref{tab:results_multi} we present the F-scores of the multi-class classification problem, where are included the five transient classes and the non-transient class using only the CNN method with images. 
Compared to the previous task, the overall performance is worse for every transient class, showing that this task is more difficult when non-transient objects are included.
The F-measure shows that the best classification is for the non-transient class with a score of 95.29. The best transient class classified correctly is the CV with a score of 54.76, followed by the SN class with a score of 46.59.

\section{Conclusions}

There is an increasing interest in automatized methods to detect transient sources. 
Some of these methods are based on Deep Learning techniques that require the use of large and realistic and data sets for its training.
Having public and easy-to-access data sets can trigger the development of new deep learning applications for transient detection.

In this paper we presented such a data set.
We named it Deep-TAO for Deep-learning Transient Astronomical Object (Deep-TAO).
This is the first public and easy-to-access data set based on real images that can be used to train and improve Deep Learning algorithms in the task of Transient classification.
The data set is a compilation of images extracted and transform from the Catalina Real-Time Transient Survey (CRTS). 
Deep-TAO includes $3,807$ transient and $12,500$ non-transient objects with a total of $1,249,079$ real astronomical images. Deep-TAO is publicly available at \url{https://github.com/MachineLearningUniandes/}.   

We demonstrated the utility of Deep-TAO with a set of deep learning experiments and comparisons against a machine learning algorithm. 
We explored the transient versus non-transient task, the fine-grained multi-classification task between five transient classes, and finally a fine-grained multi-classification task with six classes, 5 transients classes, and non-transient as another class. 

In the three tasks we used the same architecture, a Densely Connected Convolutional Network with $L=70$ layers and a growth rate $k=32$ 
motivated by the more complex architecture in \cite{Gomez2020}. 
In the fine-grained multi-classification task between five transient classes we made a comparison between a classification based on a CNN with images and the classification on light curves with a random forest with 200 trees based on the work by \cite{Neira2020}.
The results showed that CNN consistently has a better performance.

Deep-TAO is public with files in a FITS format to facilitate its usability in different projects. 
The realism of Deep-TAO provides and additional motivation to train new learning-based models to be used by next generation experiments in time-domain astronomy and hopefully it will also motivate the creation of more datasets with a similar structure: realistic, fully labeled, open and easy to access.

\section{Acknowledgements}
The authors thank the Office of the Vice Rector for Research at the Universidad de los Andes for supporting this project by the grant SPATIO TEMPORAL TRANSIENT OBJECT / P17.246622.004/01.
JFSP and JEFR acknowledge the support of the INV-2021-126-2256 and INV-2022-137-2394 projects of the Universidad de Los Andes, Facultad de Ciencias.
We also thank contributors and collaborators of the open-source packages fundamental to our work: NumPy \citep{VanDerWalt2011}, the Jupyter notebook \citep{Kluyver2016}, matplotlib \citep{Hunter2007} and pandas \citep{Mckinney2010}. 
CRTS and CSDR2 are supported by the U.S. National Science Foundation under grant NSF grants AST-1313422, AST-1413600, and AST-1518308. The CSS survey is funded by the National Aeronautics and Space Administration under Grant No. NNG05GF22G was issued through the Science Mission Directorate Near-Earth Objects Observations Program.

%%%%%%%%%%%%%%%%%%%%%%%%%%%%%%%%%%%%%%%%%%%%%%%%%%
\section*{Data Availability}

The Deep-TAO data set is publicly available at \url{https://github.com/MachineLearningUniandes/} into two different repositories, one for \href{https://github.com/MachineLearningUniandes/TAO_transients}{transients objects} and other for \href{https://github.com/MachineLearningUniandes/TAO_non-transients}{non-transients}.

\bibliography{bib}

\end{document}